\documentstyle[aps,psfig,multicol]{revtex}
\newcommand{\bcols}{\ifpreprintsty\else\begin{multicols}{2}\fi}
\newcommand{\ecols}{\ifpreprintsty\else\end{multicols}\fi}
\begin{document}
\draft
\title{Photonic excess noise and wave localization}
\author{C. W. J. Beenakker and M. Patra}
\address{Instituut-Lorentz, Universiteit Leiden,
P.O. Box 9506, 2300 RA Leiden, The Netherlands}
\author{P. W. Brouwer}
\address{Laboratory of Atomic and Solid State Physics,
Cornell University, Ithaca NY 14853, USA}
\date{January 2000}
\maketitle
\begin{abstract}
This is a theory for the effect of localization on the super-Poissonian noise
of radiation propagating through an absorbing disordered waveguide.
Localization suppresses both the mean photon current $\bar{I}$ and the noise
power $P$, but the Fano factor $P/\bar{I}$ is found to remain unaffected. For
strong absorption the Fano factor has the universal value $1+\frac{3}{2}f$
(with $f$ the Bose-Einstein function), regardless of whether the waveguide is
long or short compared to the localization length.
\end{abstract}
\bcols

The coherent radiation from a laser has Poisson statistics \cite{Wal94,Man95}.
Its noise power $P_{\rm Poisson}$ equals the mean current $\bar{I}$ (in units
of photons per second). Elastic scattering has no effect on the noise, because
the radiation remains in a coherent state. The coherent state is degraded by
absorption, resulting in an excess noise $P-P_{\rm Poisson}>0$ \cite{Hen96}.
The Fano factor $P/P_{\rm Poisson}$ deviates from unity by an amount
proportional to the Bose-Einstein function $f$. It is a small effect ($f\sim
10^{-2}$ at room temperature for infrared frequencies), but of interest because
of its fundamental origin: The excess noise is required to preserve the
canonical commutation relations of the electromagnetic field in an absorbing
dielectric \cite{Mat95,Gru95,Bar96}.

The interference of multiply scattered waves may lead to localization
\cite{She90}. Localization suppresses both the mean current and the
fluctuations --- on top of the suppression due to absorption. Localization is
readily observed in a waveguide geometry \cite{Sto99}, where it sets in once
the length $L$ of the waveguide becomes longer than the localization length
$\xi\simeq Nl$ (with $l$ the mean free path and $N$ the number of propagating
modes). Typically, $\xi$ is much larger than the absorption length $\xi_{\rm
a}$, so that localization and absorption coexist. The interplay of absorption
and localization has been studied previously for the mean current
\cite{And84,Wea93,Yos94,Bro98}. Here we go beyond these studies to include the
current fluctuations.

It is instructive to contrast the super-Poissonian photonic noise with the
sub-Poissonian electronic analogue. In the case of electrical conduction
through a disordered wire, the (zero-temperature) noise power is smaller than
the Poisson value as a result of Fermi-Dirac statistics. The reduction is a
factor $1/3$ in the absence of localization \cite{Bee92,Nag92}. The effect of
localization is to restore Poisson statistics, so that the Fano factor
increases from $1/3$ to 1 when $L$ becomes larger than $\xi$. What we will show
in this paper is that the photonic excess noise responds entirely differently
to localization: Although localization suppresses $P$ and $\bar{I}$, the Fano
factor remains unaffected, equal to the value $1+\frac{3}{2}f$ obtained in the
absence of localization \cite{Pat99,Bee99}.

\begin{figure}[tb]
\centerline{\psfig{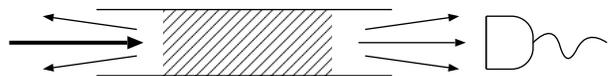}}
\medskip
\caption[]{
Monochromatic radiation (thick arrow) is incident on a disordered absorbing
medium (shaded), embedded in a waveguide. The transmitted radiation is measured
by a photodetector.
\label{fig1}}
\end{figure}

Let us begin our analysis with a more precise formulation of the problem. The
noise power
\begin{equation}
P=\int_{-\infty}^{\infty}dt\,\overline{\delta I(0)\delta I(t)} \label{Pdef}
\end{equation}
quantifies the size of the time-dependent fluctuations of the photon current
$I(t)=\bar{I}+\delta I(t)$. (The bar $\overline{\cdots}$ indicates an average
over many measurements on the same system.) For a Poisson process, the power
$P_{\rm Poisson}=\bar{I}$ equals the mean current and the Fano factor ${\cal
F}=P/P_{\rm Poisson}$ equals unity. We consider monochromatic radiation
(frequency $\omega_{0}$) incident in a single mode (labeled $m_{0}$) on a
waveguide containing a disordered medium (at temperature $T$). (See Fig.\
\ref{fig1}.) The incident radiation has Fano factor ${\cal F}_{\rm in}$. We
wish to know how the Fano factor changes as the radiation propagates through
the waveguide.

Starting point of our investigation is a formula that relates the Fano factor
to the scattering matrix of the medium \cite{Pat99},
\begin{eqnarray}
{\cal F}&=&1+[t^{\dagger}t]_{m_{0}m_{0}}({\cal F}_{\rm in}-1)\nonumber\\
&&\mbox{}+2f(\omega_{0},T)\frac{[t^{\dagger}
(\openone-rr^{\dagger}-tt^{\dagger})t]_{m_{0}m_{0}}}
{[t^{\dagger}t]_{m_{0}m_{0}}}. \label{Pexcess}
\end{eqnarray}
(We have assumed detection with quantum efficiency 1 in a narrow frequency
interval around $\omega_{0}$.) The function
$f(\omega,T)=[\exp(\hbar\omega/kT)-1]^{-1}$ is the Bose-Einstein function. The
transmission matrix $t$ and the reflection matrix $r$ are $N\times N$ matrices,
with $N$ the number of propagating modes at frequency $\omega_{0}$. The term
proportional to $f$ in Eq.\ (\ref{Pexcess}) is the excess noise. For a unitary
scattering matrix, $rr^{\dagger}+tt^{\dagger}$ equals the unit matrix
$\openone$, hence the excess noise vanishes.

In what follows we will assume that the incident radiation is in a coherent
state, so that ${\cal F}_{\rm in}=1$ and the deviation of ${\cal F}$ from unity
is entirely due to the excess noise. Since the Bose-Einstein function at room
temperature is negligibly small at optical frequencies, one would need to use
the coherent radiation from an infrared or microwave laser. Alternatively, one
could use a non-coherent source and extract the excess noise contribution by
subtracting the noise at low temperature from that at room temperature.

The absorbing disordered waveguide is characterized by four length scales: the
wavelength $\lambda$, the mean free path for scattering $l$, the absorption
length $\xi_{\rm a}$, and the localization length $\xi=(N+1)l$. We assume the
ordering of length scales $\lambda\ll l\ll\xi_{\rm a}\ll \xi$, which is the
usual situation \cite{Sto99}. We ask for the average $\langle{\cal F}\rangle$
of the Fano factor, averaged over an ensemble of waveguides with different
realizations of the disorder. For $L\gg\xi_{\rm a}$ we may neglect the matrix
$tt^{\dagger}$ with respect to $\openone$ in Eq.\ (\ref{Pexcess}), so that the
expression for the Fano factor (with ${\cal F}_{\rm in}=1$) takes the form
\begin{equation}
{\cal
F}=1+2f(1-C_{1}),\;\;C_{p}\equiv
\frac{[t^{\dagger}(rr^{\dagger})^{p}t]_{m_{0}m_{0}}} 
{[t^{\dagger}t]_{m_{0}m_{0}}}. \label{Pexcessloca}
\end{equation}
In the absence of localization, for $L\ll\xi$, one can simplify the calculation
of $\langle\cal F\rangle$ by averaging separately the numerator and denominator
in the coefficient $C_{1}$, since the sample-to-sample fluctuations are small.
This diffusive regime was studied in Refs.\ \cite{Pat99,Bee99}. Such a
simplification is no longer possible in the localized regime and we should
proceed differently.

We follow the general approach of Ref.\ \cite{Bro98}, by considering the change
in $\cal F$ upon attaching a short segment of length $\delta L$ to one end of
the waveguide. Transmission and reflection matrices are changed to leading
order in $\delta L$ according to
\begin{equation}
t\rightarrow t_{\delta L}(1+r r_{\delta L})t,\;\;r\rightarrow r_{\delta L}'+
t_{\delta L}(1+r r_{\delta L})r t_{\delta L}^{\rm T},\label{deltart}
\end{equation}
where the superscript ${\rm T}$ indicates the transpose of a matrix. (Because
of reciprocity the transmission matrix from left to right equals the transpose
of the transmission matrix from right to left.) The transmission and reflection
matrices $t_{\delta L}$, $r_{\delta L}$, $r_{\delta L}'$ of the short segment
have zero mean and variances
\begin{mathletters}
\label{deltartvar}
\begin{eqnarray}
\langle [r_{\delta L}]^{\vphantom{\ast}}_{kl} [r_{\delta
L}]^{\ast}_{mn}\rangle&=&
\langle [r'_{\delta L}]^{\vphantom{\ast}}_{kl} [r'_{\delta
L}]^{\ast}_{mn}\rangle\nonumber\\
&=& (\delta_{km}\delta_{ln}+\delta_{kn}\delta_{lm})\delta
L/\xi,\label{deltartvara}\\
\langle [t_{\delta L}]^{\vphantom{\ast}}_{kl} [t_{\delta
L}]^{\ast}_{mn}\rangle&=&
N^{-1}\delta_{km}\delta_{ln}(1-\delta L/l-\delta L/l_{\rm
a}),\label{deltartvarb}
\end{eqnarray}
\end{mathletters}
where $l_{\rm a}=2\xi_{\rm a}^{2}/l$ is the ballistic absorption length. All
covariances vanish. Substituting Eq.\ (\ref{deltart}) into Eq.\
(\ref{Pexcessloca}) and averaging we find the evolution equation
\begin{eqnarray}
\xi\frac{d\langle C_{1}\rangle}{dL}&=&-2\langle
C_{1}\rho_{1}\rangle+\langle\rho_{2}\rangle\nonumber\\
&&\mbox{}-\frac{\xi l}{\xi_{\rm a}^{2}}\langle C_{1}\rangle+1+2\langle
C_{2}-C_{1}\rangle-\langle C_{1}^{2}\rangle\nonumber\\
&&\mbox{}-4\,{\rm
Re}\,\bigl\langle[t^{\dagger}t]_{m_{0}m_{0}}^{-2}
[t^{\dagger}rt^{\ast}]_{m_{0}m_{0}}
[t^{\rm T}r^{\dagger}rr^{\dagger}t]_{m_{0}m_{0}}\bigr\rangle\nonumber\\
&&\mbox{}+2\bigl\langle(1+C_{1})[t^{\dagger}t]_{m_{0}m_{0}}^{-2}
\bigl|[t^{\dagger}rt^{\ast}]_{m_{0}m_{0}}\bigr|^{2}\bigr\rangle,
\label{C1evolution}
\end{eqnarray}
where we have defined $\rho_{p}={\rm tr}\,(1-rr^{\dagger})^{p}$.

For $L\gg\xi_{\rm a}$ we may replace the average of the product $\langle
C_{1}\rho_{1}\rangle$ by the product of averages $\langle
C_{1}\rangle\langle\rho_{1}\rangle$, because \cite{Bro98} statistical
correlations with traces that involve reflection matrices only are of relative
order $\xi_{\rm a}/\xi$ --- which we have assumed to be $\ll 1$. The moments of
the reflection matrix are given for $L\gg\xi_{\rm a}$ by \cite{Bee96}
\begin{equation}
\langle\rho_{p}\rangle=\frac{\Gamma(p-1/2)}{\sqrt{\pi}\,\Gamma(p)}\,
\frac{\xi}{\xi_{\rm a}},\label{rmoments}
\end{equation}
hence they are $\gg 1$ and also $\gg \xi l/\xi_{\rm a}^{2}$. We may therefore
neglect the terms in the second, third, and fourth line of Eq.\
(\ref{C1evolution}). What remains is the differential equation
\begin{equation}
\xi\frac{d\langle C_{1}\rangle}{dL}=-2\langle
C_{1}\rangle\langle\rho_{1}\rangle+\langle\rho_{2}\rangle,\label{C1remains}
\end{equation}
which for $L\gg\xi_{\rm a}$ has the solution
\begin{equation}
\langle
C_{1}\rangle=\frac{\langle\rho_{2}\rangle}{2\langle\rho_{1}\rangle}=\frac{1}{4}.
\label{C1solution}
\end{equation}
We conclude that the average Fano factor $\langle{\cal F}\rangle=1+2f(1-\langle
C_{1}\rangle)\rightarrow 1+\frac{3}{2}f$ for $L\gg\xi_{\rm a}$, regardless of
whether $L$ is small or large compared to $\xi$.

\begin{figure}[tb]
\centerline{\psfig{figure=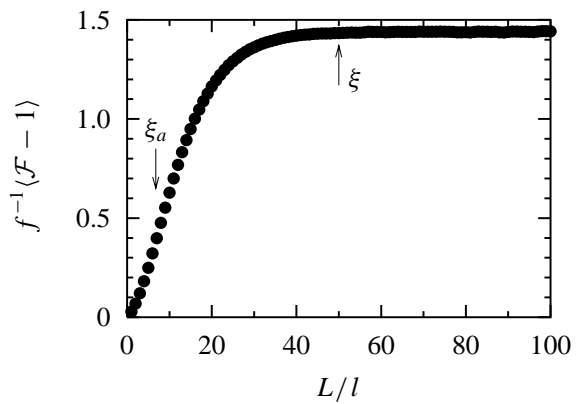,width= 8cm}}
\medskip
\caption[]{
Length dependence of the average Fano factor, computed from Eq.\
(\protect\ref{Pexcess}) with ${\cal F}_{\rm in}=1$. The data points result from
a numerical simulation for an absorbing disordered waveguide with $N=50$
propagating modes. Arrows indicate the absorption length $\xi_{\rm a}$ and the
localization length $\xi$. The average Fano factor is not affected by
localization.
\label{fig2}}
\end{figure}

\begin{figure}[tb]
\centerline{\psfig{figure=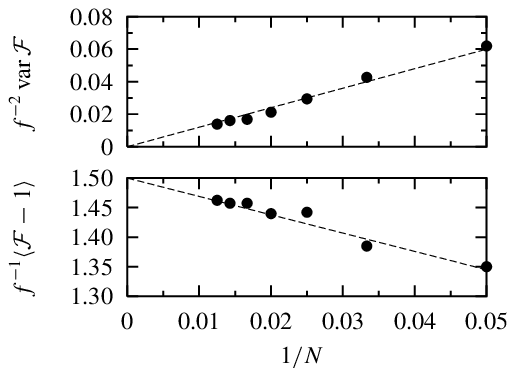,width= 8cm}}
\medskip
\caption[]{
Dependence of the average and variance of the Fano factor on the number $N$ of
propagating modes, for fixed length $L=260\,l=38.5\,\xi_{\rm a}$ of the
waveguide. The length is larger than the localization length $\xi=(N+1)l$ for
all data points. The dashed lines extrapolate to the theoretical expectation
for $1/N\rightarrow 0$.
\label{fig3}}
\end{figure}

To support this analytical calculation we have carried out numerical
simulations. The absorbing disordered waveguide is modeled by a two-dimensional
square lattice (lattice constant $a$). The dielectric constant $\varepsilon$
has a real part that fluctuates from site to site and a non-fluctuating
imaginary part. The multiple scattering of a scalar wave $\Psi$ is described by
discretizing the Helmholtz equation $[\nabla^{2}+
(\omega_{0}/c)^{2}\varepsilon]\Psi=0$ and computing the transmission and
reflection matrices using the recursive Green function technique \cite{Bar91}.
The mean free path $l=20\,a$ and the absorption length $\xi_{\rm a}=135\,a$ are
determined from the average transmission probability $N^{-1}\langle{\rm
tr}\,tt^{\dagger}\rangle=l/\xi_{\rm a}\sinh(L/\xi_{\rm a})$ in the diffusive
regime \cite{Bro98}. Averages were performed over the $N/2$ modes $m_{0}$ near
normal incidence and over some $10^{2}-10^{3}$ realizations of the disorder.
Results are shown in Figs.\ \ref{fig2} and \ref{fig3}.

The length dependence of the average Fano factor is plotted in Fig.\ 2, for
$N=50$ and $L$ ranging from $0$ to $2\xi$. Clearly, localization has no effect.
The limiting value of $f^{-1}\langle{\cal F}-1\rangle$ resulting from this
simulation is slightly smaller than the value $3/2$ predicted by the analytical
theory for $N\gg 1$. The $N$-dependence of $\langle{\cal F}\rangle$ in the
localized regime is shown in Fig.\ 3. A line through the data points
extrapolates to the theoretical expectation $f^{-1}\langle{\cal
F}-1\rangle\rightarrow 3/2$ for $N\rightarrow\infty$. Fig.\ 3 also shows the
variance of the Fano factor. The variance extrapolates to $0$ for
$N\rightarrow\infty$, indicating that ${\cal F}=P/\bar{I}$ becomes
self-averaging for large $N$. This is in contrast to $P$ and $\bar{I}$
themselves, which fluctuate strongly in the localized regime.

In conclusion, we have demonstrated that localization of radiation in an
absorbing disordered waveguide has no effect on the ratio of the excess noise
and the mean current. In the limit of a large number of propagating modes, this
ratio is self-averaging and takes on the universal value of $3/2$ times the
Bose-Einstein function. Observation of this photonic analogue of the universal
$1/3$ reduction of electronic shot noise presents an experimental challenge.

This work was supported by the Dutch Science Foundation NWO/FOM.

\ecols

\end{document}